\documentclass[showpacs,column,showkeys,amsmath,amssymb,superscriptaddress,aps,prl]{revtex4}

\begin{document}
\title{{\bf Solution of the }$D${\bf -dimensional Klein-Gordon equation with equal
scalar and vector ring-shaped pseudoharmonic potential }}
\author{Sameer M. Ikhdair\thanks{%
sikhdair@neu.edu.tr} and \ Ramazan Sever\thanks{%
sever@metu.edu.tr}}
\address{$^{\ast }$Department of Physics, \ Near East University, Nicosia, Cyprus,
Mersin 10, Turkey\\
$^{\dagger }$Department of Physics, Middle East Technical University, 06531
Ankara, Turkey.}
\date{\today}

\begin{abstract}
We present the exact solution of the Klein-Gordon equation in $D$-dimensions
in the presence of the noncentral equal scalar and vector pseudoharmonic
potential plus the new ring-shaped potential using the Nikiforov-Uvarov
method. We obtain the exact bound-state energy levels and the corresponding
eigen functions for a spin-zero particles. We also find that the solution
for this noncentral ring-shaped pseudoharmonic potential can be reduced to
the three-dimensional pseudoharmonic solution once the coupling constant of
the noncentral part of the potential becomes zero.

Keywords: Bound-states, energy eigenvalues and eigenfunctions, Klein-Gordon
equation, pseudoharmonic potential, ring-shaped potential, non-central
potentials, Nikiforov and Uvarov method.

PACS numbers: 03.65.-w; 03.65.Fd; 03.65.Ge.
\end{abstract}
\maketitle

\section{Introduction}

\noindent In nuclear and high energy physics \cite{1,2}, one of the interesting
problems is to obtain exact solutions of the relativistic wave equations
like Klein-Gordon, Dirac and Salpeter wave equations for mixed vector and
scalar potential. The Klein-Gordon and Dirac wave equations are frequently
used to describe the particle dynamics in relativistic quantum mechanics.
The Klein-Gordon equation has also been used to understand the motion of a
spin-$0$ particle in large class of potentials. In recent years, much works
have been done to solve these relativistic wave equations for various
potentials by using different methods. These relativistic equations contain
two objects: the four-vector linear momentum operator and the scalar rest
mass. They allow us to introduce two types of potential coupling, which are
the four-vector potential (V) and the space-time scalar potential (S).

For the case $S=\pm V,$ the solution of these wave equations with physical
potentials has been studied recently. The exact solutions of these equations
are possible only for certain central potentials such as Morse potential
\cite{3}, Hulth\'{e}n potential \cite{4}, Woods-Saxon potential \cite{5}, P\"{o}schl-Teller
potential \cite{6}, reflectionless-type potential \cite{7}, pseudoharmonic oscillator
\cite{8}, ring-shaped harmonic oscillator \cite{9}, $V_{0}\tanh ^{2}(r/r_{0})$
potential \cite{10}, five-parameter exponential potential \cite{11}, Rosen-Morse
potential \cite{12}, and generalized symmetrical double-well potential \cite{13}, etc
by using different methods. It is remarkable that in most works in this
area, the scalar and vector potentials are almost taken to be equal (i.e., $%
S=V$) \cite{2,14}. However, in some few other cases, it is considered the case
where the scalar potential is greater than the vector potential (in order to
guarantee the existence of Klein-Gordon bound states) (i.e., $S>V$) \cite{15,16,17,18,19}.
Nonetheless, such physical potentials are very few. The bound-state
solutions for the last case is obtained for the exponential potential for
the $s$-wave Klein-Gordon equation when the scalar potential is greater than
the vector potential \cite{15}.

On the other hand, the other exactly solvable potentials are the noncentral
ring-shaped potentials \cite{20}. These potentials involve an attractive Coulomb
potential plus a repulsive inverse square potential, that is, one like the
Coulombic ring-shaped potential \cite{21,22} revived in quantum chemistry by
Hartmann {\it et al } \cite{23}. The oscillatory ring-shaped potential studied by
Quesne \cite{24} have been investigated using various quantum mechanical
approaches \cite{25}. In taking the relativistic effects into account for spin-$0$
particle in the presence of a class of noncentral potentials, Yasuk {\it et
al} \cite{26} applied the Nikiforov-Uvarov method \cite{27} to solve the Klein-Gordon
equation for the noncentral Coulombic ring-shaped potential \cite{21} for the
case $V=S.$ Further, Berkdemir \cite{28} also applied the same method to solve
the Klein-Gordon equation for the Kratzer-type potential.

Recently, Chen and Dong \cite{29} proposed a new ring-shaped potential and
obtained the exact solution of the Schr\"{o}dinger equation for the Coulomb
potential plus this new ring-shaped potential which has possible
applications to ring-shaped organic molecules like cyclic polyenes and
benzene. This type of potential used by Ref. \cite{29} appears to be very similar
to the potential used by Ref. \cite{26}. Additionally, Cheng and Dai \cite{30},
proposed a new potential consisting from the modified Kratzer's potential
\cite{31} plus the new proposed ring-shaped potential in \cite{29}. They have
presented the energy eigenvalues for this proposed exactly-solvable
non-central potential in three dimensional Schr\"{o}dinger equation using
the NU method. The two quantum systems solved by Refs \cite{29,30} are closely
relevant to each other as they deal with a Coulombic field interaction
except for an additional change in the angular momentum barrier. In
addition, the $D$-dimensional Schr\"{o}dinger and Klein-Gordon wave
equations have been solved for some types of ring-shaped potentials using
the NU method \cite{22,25,32}.

The aim of the present paper is to obtain the exact bound-state solutions of
the $D$-dimensional Klein-Gordon with an oscillatory-type plus new
ring-shaped potential. The radial and angular parts of the Klein-Gordon
equation with this type of noncentral potential are solved using the NU
method.

This work is organized as follows: in section \ref{TKG}, we shall present
the Klein-Gordon equation in spherical coordinates for spin-$0$ particle
with an equal scalar and vector oscillatory-type ring-shaped potential. We
separate the wave equation into radial and angular parts. Section \ref{NUM}
is devoted to a brief description of the NU method. In section \ref{ES}, we
present the exact bound-state solutions to the radial and angular equations
in $D$-dimensions. Finally, the relevant conclusions are given in section
\ref{C}.

\section{The Klein-Gordon Equation with Equal Scalar and Vector Potentials}

\label{TKG}In relativistic quantum mechanics, we usually use the
Klein-Gordon equation for describing a scalar particle, i.e., the spin-$0$
particle dynamics. The discussion of the relativistic behavior of spin-zero
particles requires understanding the single particle spectrum and the exact
solutions to the Klein Gordon equation which are constructed by using the
four-vector potential ${\bf A}_{\lambda }$ $(\lambda =0,1,2,3)$ and the
scalar potential $(S)$. In order to simplify the solution of the
Klein-Gordon equation, the four-vector potential can be written as ${\bf A}%
_{\lambda }=(A_{0},0,0,0).$ The first component of the four-vector potential
is represented by a vector potential $(V),$ i.e., $A_{0}=V.$ In this case,
the motion of a relativistic spin-$0$ particle in a potential is described
by the Klein-Gordon equation with the potentials $V$ and $S$ \cite{1}$.$ For the
case $S\geq V,$ there exist bound-state (real) solutions for a relativistic
spin-zero particle \cite{15,16,17,18,19}. On the other hand, for $S=V,$ the Klein-Gordon
equation reduces to a Schr\"{o}dinger-like equation and thereby the
bound-state solutions are easily obtained by using the well-known methods
developed in nonrelativistic quantum mechanics \cite{2}.

The Klein-Gordon equation describing a scalar particle (spin-$0$ particle)
with scalar $S(r,\theta ,\varphi )$ and vector $V(r,\theta ,\varphi )$
potentials is given by \cite {2,14}

\begin{equation}
\left\{ {\bf P}^{2}-\left[ V(r,\theta ,\varphi )-E_{R}\right] ^{2}+\left[
S(r,\theta ,\varphi )+\mu \right] ^{2}\right\} \psi (r,\theta ,\varphi )=0,
\end{equation}
where $\mu $ is the rest mass, $E_{R}$ is the relativistic energy, ${\bf P}$
is the momentum operator and $S$ and $V$ are the scalar and vectorial
potentials. Alhaidari {\it et al } \cite{14} concluded that only the choice $S=V$
produces a non trivial nonrelativistic limit with a potential function $2V$
and not $V.$ Accordingly, it would be natural to scale the potential terms
in Eq. (1) so that in the nonrelativistic limit the interaction potential
becomes $V$ and not $2V.$

Therefore, they modified Eq. (1) to read as follows (in the relativistic
atomic units $\hbar =c=1$):
\begin{equation}
\left\{ {\bf \nabla }^{2}+\left[ \frac{1}{2}V(r,\theta ,\varphi )-E_{R}%
\right] ^{2}-\left[ \frac{1}{2}S(r,\theta ,\varphi )+\mu \right]
^{2}\right\} \psi (r,\theta ,\varphi )=0.
\end{equation}
After substituting $S(r,\theta ,\varphi )=V(r,\theta ,\varphi ),$ the equal
scalar and vector potentials case, Eq. (2) becomes
\begin{equation}
\left\{ {\bf \nabla }^{2}-(E_{R}+\mu )V(r,\theta ,\varphi )+E_{R}^{2}-\mu
^{2}\right\} \psi (r,\theta ,\varphi )=0,
\end{equation}
If we take the interaction potential in Eq. (3) as a general noncentral
oscillatory-type ring-shaped potential, the $D$-dimensional Klein-Gordon
equation is separated into variables and the equation can be solved through
the NU method.

We take the interaction potential in Eq. (3) to be of an oscillatory-type
plus new ring-shaped potential which is the potential of a diatomic
molecule, \cite{32}:
\[
V(r,\theta ,\varphi )=V_{1}(r)+\frac{V_{2}(\theta )}{r^{2}}+\frac{%
V_{3}(\varphi )}{r^{2}\sin ^{2}\theta },
\]
\begin{equation}
V_{1}(r)=Ar^{2}+\frac{B}{r^{2}}+C,\text{ }V_{2}(\theta )=\beta ctg^{2}\theta
,\text{ }V_{3}(\varphi )=0,
\end{equation}
where $A=a_{0}r_{0}^{-2},$ $B=a_{0}r_{0}^{2},$ $C=-2a_{0}$ and $\beta $ is
positive real constant with $a_{0}$ is the dissociation energy and $r_{0}$
is the equilibrium internuclear distance \cite{33,34}. The potentials in Eq. (4)
can be reduced to pseudoharmonic potential in the limiting case of $\beta =0$
\cite{34}$.$ Nonetheless, the energy spectrum for this potential can be obtained
directly by considering it as special case of the general non-central
seperable potentials \cite{20}.

Our aim is to derive analytically the exact energy spectrum for a moving
particle in the presence of a non central potential given by Eq. (4) in a
very simple way. We begin by considering the Schr\"{o}dinger equation in
arbitrary dimensions $D$ for our proposed potential \cite{32}

\[
\left\{ \nabla _{D}^{2}+\frac{2\mu }{\hbar ^{2}}\left[ E-V(r)-\frac{1}{r^{2}}%
V(\theta )\right] \right\} \psi _{\ell _{1}\cdots \ell _{D-2}}^{(\ell
_{D-1}=\ell )}({\bf x})=0,
\]
\[
\nabla _{D}^{2}=\frac{\partial ^{2}}{\partial r^{2}}+\frac{(D-1)}{r}\frac{%
\partial }{\partial r}
\]
\[
+\frac{1}{r^{2}}\left[ \frac{1}{\sin ^{D-2}\theta _{D-1}}\frac{\partial }{%
\partial \theta _{D-1}}\left( \sin ^{D-2}\theta _{D-1}\frac{\partial }{%
\partial \theta _{D-1}}\right) -\frac{L_{D-2}^{2}}{\sin ^{2}\theta _{D-1}}%
\right] ,
\]
\begin{equation}
\psi _{\ell _{1}\cdots \ell _{D-2}}^{(\ell )}({\bf x})=R_{\ell }(r)Y_{\ell
_{1}\cdots \ell _{D-2}}^{(\ell )}(\widehat{{\bf x}}),\text{ }R_{\ell
}(r)=r^{-(D-1)/2}g(r),\text{ }
\end{equation}
where $\mu $ $\ $and $E$ denote the reduced mass and energy of two
interacting particles, respectively. ${\bf x}$ is a $D$-dimensional position
vector with the hyperspherical Cartesian components $x_{1},x_{2},\cdots
,x_{D}$ given as follows\footnote{%
It is worth noting that such a definition was introduced by Erd\'{e}lyi
early in 1950s (cf. \cite{40}, pp.232-5, chapter 11) even though the notation
used by him is quite different from that by Louck and Chatterjee.} \cite{35-40}:

\[
x_{1}=r\cos \theta _{1}\sin \theta _{2}\cdots \sin \theta _{D-1},
\]
\[
x_{2}=r\sin \theta _{1}\sin \theta _{2}\cdots \sin \theta _{D-1},
\]
\[
x_{3}=r\cos \theta _{2}\sin \theta _{3}\cdots \sin \theta _{D-1},
\]
\[
\vdots
\]
\[
x_{j}=r\cos \theta _{j-1}\sin \theta _{j}\cdots \sin \theta _{D-1},\text{ }%
3\leq j\leq D-1,
\]
\[
\vdots
\]
\[
x_{D-1}=r\cos \theta _{D-2}\sin \theta _{D-1},
\]
\begin{equation}
x_{D}=r\cos \theta _{D-1},\text{ }\sum\limits_{j=1}^{D}x_{j}^{2}=r^{2},
\end{equation}
for $D=2,3,\cdots .$ We have $x_{1}=r\cos \varphi ,$ $x_{2}=r\sin \varphi $
for $D=2$ and $x_{1}=r\cos \varphi \sin \theta ,$ $x_{2}=r\sin \varphi \sin
\theta ,$ $x_{3}=r\cos \theta $ for $D=3.$ The Laplace operator $\nabla
_{D}^{2}$ is defined by
\begin{equation}
\nabla _{D}^{2}=\sum\limits_{j=1}^{D}\frac{\partial ^{2}}{\partial x_{j}^{2}}%
.
\end{equation}
The volume element of the configuration space is given by
\begin{equation}
\prod\limits_{j=1}^{D}dx_{j}=r^{D-1}drd\Omega ,\text{ }d\Omega
=\prod\limits_{j=1}^{D-1}(\sin \theta _{j})^{j-1}d\theta _{j},
\end{equation}
where $r\in \lbrack 0,\infty ),$ $\theta _{1}\in \lbrack 0,2\pi ]$ and $%
\theta _{j}\in \lbrack 0,\pi ],$ $j\in \lbrack 2,D-1].$ The wave function $%
\psi _{\ell _{1}\cdots \ell _{D-2}}^{(\ell )}({\bf x)}$ with a given angular
momentum $\ell $ can be decomposed as a product of a radial wave function $%
R_{\ell }(r)$ and the generalized spherical harmonics $Y_{\ell _{1}\cdots
\ell _{D-2}}^{(\ell )}(\widehat{{\bf x}})$ as \cite{35}

\[
\psi _{\ell _{1}\cdots \ell _{D-2}}^{(\ell )}({\bf x})=R_{\ell }(r)Y_{\ell
_{1}\cdots \ell _{D-2}}^{(\ell )}(\widehat{{\bf x}}),\text{ }
\]
\[
Y_{\ell _{1}\cdots \ell _{D-2}}^{(\ell )}(\widehat{{\bf x}})=Y(\ell
_{1},\ell _{2},\cdots ,\ell _{D-2},\ell ),\text{ }\ell =\left| m\right|
\text{ for }D=2,
\]
\[
R_{\ell }(r)=r^{-(D-1)/2}g(r),\text{ }
\]
\begin{equation}
Y_{\ell _{1}\cdots \ell _{D-2}}^{(\ell )}(\widehat{{\bf x}}=\theta
_{1},\theta _{2},\cdots ,\theta _{D-1})=\Phi (\theta _{1}=\varphi )H(\theta
_{2},\cdots ,\theta _{D-1}),
\end{equation}
which is the simultaneous eigenfunction of $L_{j}^{2}:$%
\[
L_{1}^{2}Y_{\ell _{1}\cdots \ell _{D-2}}^{(\ell )}(\widehat{{\bf x}}%
)=m^{2}Y_{\ell _{1}\cdots \ell _{D-2}}^{(\ell )}(\widehat{{\bf x}}),
\]
\[
L_{j}^{2}Y_{\ell _{1}\cdots \ell _{D-2}}^{(\ell )}(\widehat{{\bf x}})=\ell
_{j}(\ell _{j}+j-1)Y_{\ell _{1}\cdots \ell _{D-2}}^{(\ell )}(\widehat{{\bf x}%
}),
\]
\[
\ell =0,1,\cdots ,\ell _{k}=0,1,\cdots ,\ell _{k+1},\text{ }j\in \lbrack
1,D-1],\text{ }k\in \lbrack 2,D-2],
\]
\[
\ell _{1}=-\ell _{2},-\ell _{2}+1,\cdots ,\ell _{2}-1,\ell _{2},
\]
\begin{equation}
L_{D-1}^{2}Y_{\ell _{1}\cdots \ell _{D-2}}^{(\ell )}(\widehat{{\bf x}})=\ell
(\ell +D-2)Y_{\ell _{1}\cdots \ell _{D-2}}^{(\ell )}(\widehat{{\bf x}}),
\end{equation}
where the unit vector along ${\bf x}$ is usually denoted by $\widehat{{\bf x}%
}={\bf x}/r{\bf .}$

Hence for a nonrelativistic treatment with the potential given in Eq. (4),
the Schr\"{o}dinger equation in spherical coordinates is

\[
\left\{ \frac{1}{r^{D-1}}\frac{\partial }{\partial r}\left( r^{D-1}\frac{%
\partial }{\partial r}\right) -\frac{\ell _{D-1}(\ell _{D-1}+D-2)}{r^{2}}+%
\frac{2\mu }{\hbar ^{2}}\left( E_{NR}-V_{1}(r)-\frac{V_{2}(\theta )}{r^{2}}-%
\frac{V_{3}(\varphi )}{r^{2}\sin ^{2}\theta }\right) \right\}
\]
\begin{equation}
\times R_{\ell }(r)=0,
\end{equation}
where $\mu $ and $E_{NR}$ are the reduced mass and the nonrelativistic
energy, respectively. The angular momentum operators $L_{j}^{2}$ are defined
as \cite{35,36,37,38,39}$:$%
\[
L_{1}^{2}=-\frac{\partial ^{2}}{\partial \theta _{1}^{2}},
\]
\[
L_{k}^{2}=\sum\limits_{a<b=2}^{k+1}L_{ab}^{2}=-\frac{1}{\sin ^{k-1}\theta
_{k}}\frac{\partial }{\partial \theta _{k}}\left( \sin ^{k-1}\theta _{k}%
\frac{\partial }{\partial \theta _{k}}\right) +\frac{L_{k-1}^{2}}{\sin
^{2}\theta _{k}},\text{ }2\leq k\leq D-1,
\]
\begin{equation}
L_{ab}=-i\left[ x_{a}\frac{\partial }{\partial x_{b}}-x_{b}\frac{\partial }{%
\partial x_{a}}\right] .
\end{equation}
Making use of Eqs. (10) and (12), leads to the separation of Eq. (11) into
the following set of second-order differential equations:
\begin{equation}
\frac{d^{2}\Phi (\theta _{1}=\varphi )}{d\theta _{1}^{2}}+m^{2}\Phi (\theta
_{1}=\varphi )=0,
\end{equation}

\[
\left[ \frac{1}{\sin ^{j-1}\theta _{j}}\frac{d}{d\theta _{j}}\left( \sin
^{j-1}\theta _{j}\frac{d}{d\theta _{j}}\right) +\ell _{j}(\ell
_{j}+j-1)\right.
\]
\begin{equation}
-\left. \frac{\ell _{j-1}(\ell _{j-1}+j-2)}{\sin ^{2}\theta _{j}}\right]
H(\theta _{j})=0,\text{ }j\in \lbrack 2,D-2],
\end{equation}
\[
\left[ \frac{1}{\sin ^{D-2}\theta _{D-1}}\frac{d}{d\theta _{D-1}}\left( \sin
^{D-2}\theta _{D-1}\frac{d}{d\theta _{D-1}}\right) +\lambda _{\ell }\right.
\]
\begin{equation}
-\left. \frac{1}{\sin ^{2}\theta _{D-1}}\left( L_{D-2}^{2}+\frac{2\mu C}{%
\hbar ^{2}}\cos ^{2}\theta _{D-1}\right) \right] H(\theta _{D-1})=0,
\end{equation}
\begin{equation}
\left[ \frac{1}{r^{D-1}}\frac{d}{dr}\left( r^{D-1}\frac{d}{dr}\right) -\frac{%
\lambda _{\ell }}{r^{2}}\right] R_{\ell }(r)+\frac{2\mu }{\hbar ^{2}}\left[
E_{NR}+\frac{A}{r}-\frac{B}{r^{2}}\right] R(r)=0,
\end{equation}
with $m^{2}$ and $\lambda _{\ell }=\ell (\ell +D-2)$ are two separation
constants whereas $\mu $ and $E_{NR}$ are the reduced mass and the
nonrelativistic energy, respectively.

On the other hand, in the relativistic atomic units ($\hbar =c=1$), the $D$%
-dimensional Klein-Gordon equation in Eq. (1) becomes \cite{28}

\[
\left\{ \frac{1}{r^{D-1}}\frac{\partial }{\partial r}\left( r^{D-1}\frac{%
\partial }{\partial r}\right) -\frac{\widetilde{\ell }(\widetilde{\ell }+D-2)%
}{r^{2}}\right.
\]

\begin{equation}
-\left. \left( E_{R}+\mu \right) \left( V_{1}(r)+\frac{V_{2}(\theta )}{r^{2}}%
\right) +\left( E_{R}^{2}-\mu ^{2}\right) \right\} \psi _{n\widetilde{\ell }%
m}(r,\theta ,\varphi )=0.
\end{equation}
With the total wave function has the same representation as in Eq. (9) but
with the transformation $\ell \rightarrow \widetilde{\ell },$

\begin{equation}
\psi _{n\widetilde{\ell }m}(r,\theta ,\varphi )=R_{\widetilde{\ell }}(r)Y_{%
\widetilde{\ell }}^{m}(\theta ,\varphi ),\text{ }R_{\widetilde{\ell }%
}(r)=r^{-(D-1)/2}g(r),\text{ }Y_{\widetilde{\ell }}^{m}(\theta ,\varphi
)=\prod\limits_{j=2}^{D-1}H_{j}(\theta )\Phi (\varphi ).
\end{equation}
and employing the method of separation of variables leads to the following
differential equations \cite{25,32,41}:

\begin{equation}
\frac{d^{2}\Phi (\varphi )}{d\varphi ^{2}}+m^{2}\Phi (\varphi )=0,
\end{equation}
\[
\left[ \frac{1}{\sin ^{j-1}\theta _{j}}\frac{d}{d\theta _{j}}\left( \sin
^{j-1}\theta _{j}\frac{d}{d\theta _{j}}\right) +\widetilde{\ell }_{j}(%
\widetilde{\ell }_{j}+j-1)\right.
\]
\begin{equation}
-\left. \frac{\widetilde{\ell }_{j-1}(\widetilde{\ell }_{j-1}+j-2)}{\sin
^{2}\theta _{j}}\right] H(\theta _{j})=0,\text{ }j\in \lbrack 2,D-2],
\end{equation}
\[
\left[ \frac{1}{\sin ^{D-2}\theta _{D-1}}\frac{d}{d\theta _{D-1}}\left( \sin
^{D-2}\theta _{D-1}\frac{d}{d\theta _{D-1}}\right) +\lambda _{\widetilde{%
\ell }}\right.
\]
\begin{equation}
-\left. \frac{\widetilde{L}_{D-2}^{2}+C\alpha _{2}^{2}\cos ^{2}\theta _{D-1}%
}{\sin ^{2}\theta _{D-1}}\right] H(\theta _{D-1})=0,
\end{equation}

\begin{equation}
\frac{1}{r^{D-1}}\frac{d}{dr}\left( r^{D-1}\frac{dR(r)}{dr}\right) -\left[
\frac{\lambda _{\widetilde{\ell }}}{r^{2}}+\alpha _{2}^{2}\left( \alpha
_{1}^{2}-\frac{A}{r}+\frac{B}{r^{2}}\right) \right] R_{\widetilde{\ell }%
}(r)=0,
\end{equation}
where $\alpha _{1}^{2}=\mu -E_{R},$ $\alpha _{2}^{2}=\mu +E_{R},$ $m$ and $%
\widetilde{\ell }$ are constants with $m^{2}$ and $\lambda _{\widetilde{\ell
}}=\widetilde{\ell }(\widetilde{\ell }+D-2)$ are the separation constants.

Equations (19)-(22) have the same functional form as Eqs (13)-(16).
Therefore, the solution of the Klein-Gordon equation can be reduced to the
solution of the Schr\"{o}dinger equation with the appropriate choice of
parameters: $\widetilde{\ell }\rightarrow \ell ,$ $\alpha
_{1}^{2}\rightarrow -E_{NR\text{ }}$ and $\alpha _{2}^{2}\rightarrow 2\mu
/\hbar ^{2}.$

The solution of Eq. (19) is well-known periodic and must satisfy the period
boundary condition $\Phi (\varphi +2\pi )=\Phi (\varphi )$ which is the
azimuthal angle solution:
\begin{equation}
\Phi _{m}(\varphi )=\frac{1}{\sqrt{2\pi }}\exp (\pm im\varphi ),\text{ \ }%
m=0,1,2,.....
\end{equation}

Additionally, Eqs (20)-(22), the polar angle and radial equations, are to be
solved by using NU method \cite{27} which is given briefly in the following
section.

\section{Nikiforov-Uvarov Method}

\label{NUM}The NU method is based on reducing the second-order differential
equation to a generalized equation of hypergeometric type \cite{20,27,43,44,45}. In
this sense, the Schr\"{o}dinger equation, after employing an appropriate
coordinate transformation $s=s(r),$ transforms to the following form:
\begin{equation}
\psi _{n}^{\prime \prime }(s)+\frac{\widetilde{\tau }(s)}{\sigma (s)}\psi
_{n}^{\prime }(s)+\frac{\widetilde{\sigma }(s)}{\sigma ^{2}(s)}\psi
_{n}(s)=0,
\end{equation}
where $\sigma (s)$ and $\widetilde{\sigma }(s)$ are polynomials, at most of
second-degree, and $\widetilde{\tau }(s)$ is a first-degree polynomial.
Using a wave function, $\psi _{n}(s),$ of \ the simple form

\begin{equation}
\psi _{n}(s)=\phi _{n}(s)y_{n}(s),
\end{equation}
reduces (24) into an equation of a hypergeometric type

\begin{equation}
\sigma (s)y_{n}^{\prime \prime }(s)+\tau (s)y_{n}^{\prime }(s)+\lambda
y_{n}(s)=0,
\end{equation}
where

\begin{equation}
\sigma (s)=\pi (s)\frac{\phi (s)}{\phi ^{\prime }(s)},
\end{equation}

\begin{equation}
\tau (s)=\widetilde{\tau }(s)+2\pi (s),\text{ }\tau ^{\prime }(s)<0,
\end{equation}
and $\lambda $ is a parameter defined as
\begin{equation}
\lambda =\lambda _{n}=-n\tau ^{\prime }(s)-\frac{n\left( n-1\right) }{2}%
\sigma ^{\prime \prime }(s),\text{ \ \ \ \ \ \ }n=0,1,2,....
\end{equation}
The polynomial $\tau (s)$ with the parameter $s$ and prime factors show the
differentials at first degree be negative. It is worthwhile to note that $%
\lambda $ or $\lambda _{n}$ are obtained from a particular solution of the
form $y(s)=y_{n}(s)$ which is a polynomial of degree $n.$ Further, the other
part $y_{n}(s)$ of the wave function (25) is the hypergeometric-type
function whose polynomial solutions are given by Rodrigues relation

\begin{equation}
y_{n}(s)=\frac{B_{n}}{\rho (s)}\frac{d^{n}}{ds^{n}}\left[ \sigma ^{n}(s)\rho
(s)\right] ,
\end{equation}
where $B_{n}$ is the normalization constant and the weight function $\rho
(s) $ must satisfy the condition \cite{27}

\begin{equation}
\frac{d}{ds}w(s)=\frac{\tau (s)}{\sigma (s)}w(s),\text{ }w(s)=\sigma (s)\rho
(s).
\end{equation}
The function $\pi $ and the parameter $\lambda $ are defined as

\begin{equation}
\pi (s)=\frac{\sigma ^{\prime }(s)-\widetilde{\tau }(s)}{2}\pm \sqrt{\left(
\frac{\sigma ^{\prime }(s)-\widetilde{\tau }(s)}{2}\right) ^{2}-\widetilde{%
\sigma }(s)+k\sigma (s)},
\end{equation}
\begin{equation}
\lambda =k+\pi ^{\prime }(s).
\end{equation}
In principle, since $\pi (s)$ has to be a polynomial of degree at most one,
the expression under the square root sign in (32) can be arranged to be the
square of a polynomial of first degree \cite{27}. This is possible only if its
discriminant is zero. In this case, an equation for $k$ is obtained. After
solving this equation, the obtained values of $k$ are substituted in (32).
In addition, by comparing equations (29) and (33), we obtain the energy
eigenvalues.

\section{Exact Solutions of the Radial and Angle-Dependent Equations}

\label{ES}

\subsection{The solutions of the $D$-dimensional angular equations}

At the beginning, we rewrite Eqs. (20) and (21) representing the angular
wave equations in the following simple forms \cite{42} :
\begin{equation}
\frac{d^{2}H(\theta _{j})}{d\theta _{j}^{2}}+(j-1)ctg\theta _{j}\frac{%
dH(\theta _{j})}{d\theta _{j}}+\left( \Lambda _{j}-\frac{\Lambda _{j-1}}{%
\sin ^{2}\theta _{j}}\right) H(\theta _{j})=0,\text{ }j\in \lbrack
2,D-2],D>3,
\end{equation}
\[
\frac{d^{2}H(\theta _{D-1})}{d\theta _{D-1}^{2}}+(D-2)ctg\theta _{D-1}\frac{%
dH(\theta _{D-1})}{d\theta _{D-1}}
\]
\begin{equation}
+\left[ \widetilde{\ell }(\widetilde{\ell }+D-2)-\frac{\Lambda _{D-2}+\beta
\alpha _{2}^{2}\cos ^{2}\theta _{D-1}}{\sin ^{2}\theta _{D-1}}\right]
H(\theta _{D-1})=0,
\end{equation}
where $\Lambda _{p}=\widetilde{\ell }_{p}(\widetilde{\ell }_{p}+p-1),$ $%
p=j-1,$ $j$ which is well-known in three-dimensional space \footnote{$\Lambda _{D-2}=m^{2}$ for $D=3$}. Equations (34) and (35) will be solved
in the following subsection. Employing $s=\cos \theta _{j},$ we transform
Eq. (34) to the associated-Legendre equation

\begin{equation}
\frac{d^{2}H(s)}{ds^{2}}-\frac{js}{1-s^{2}}\frac{dH(s)}{ds}+\frac{\Lambda
_{j}-\Lambda _{j-1}-\Lambda _{j}s^{2}}{(1-s^{2})^{2}}H(s)=0,\text{ }j\in
\lbrack 2,D-2],D>3.
\end{equation}
By comparing Eqs. (36) and (24), the corresponding polynomials are obtained
\begin{equation}
\widetilde{\tau }(s)=-js,\ \ \ \sigma (s)=1-s^{2},\ \ \widetilde{\sigma }%
(s)=-\Lambda _{j}s^{2}+\Lambda _{j}-\Lambda _{j-1}.
\end{equation}
Inserting the above expressions into Eq. (32) and taking $\sigma ^{\prime
}(s)=-2s$, one obtains the following function:

\begin{equation}
\pi (s)=\frac{(j-2)}{2}s\pm \sqrt{\left[ \left( \frac{j-2}{2}\right)
^{2}+\Lambda _{j}-k\right] s^{2}+k-\Lambda _{j}+\Lambda _{j-1}}.
\end{equation}
Following the method, the polynomial $\pi (s)$ is found to have the
following four possible values:
\begin{equation}
\pi (s)=\left\{
\begin{array}{cc}
\left( \frac{j-2}{2}+\widetilde{\Lambda }_{j-1}\right) s & \text{\ for }%
k_{1}=\Lambda _{j}-\Lambda _{j-1}, \\
\left( \frac{j-2}{2}-\widetilde{\Lambda }_{j-1}\right) s & \text{\ for }%
k_{1}=\Lambda _{j}-\Lambda _{j-1}, \\
\frac{(j-2)}{2}s+\widetilde{\Lambda }_{j-1} & \text{\ for }k_{2}=\Lambda
_{j}+\left( \frac{j-2}{2}\right) ^{2}, \\
\frac{(j-2)}{2}s-\widetilde{\Lambda }_{j-1} & \text{\ for }k_{2}=\Lambda
_{j}+\left( \frac{j-2}{2}\right) ^{2},
\end{array}
\right.
\end{equation}
where $\widetilde{\Lambda }_{p}=\widetilde{\ell }_{p}+(p-1)/2,$ with $p=j-1,$
$j$ and $j\in \lbrack 2,D-2],$ $D>3\ .$ Imposing the condition $\tau
^{\prime }(s)<0$ for Eq. (28), one selects the following physically valid
solutions with $\tau ^{\prime }=\tau ^{\prime }(\widetilde{\ell }_{j-1});$
that is, a function of the angular momentum$:$%
\begin{equation}
k_{1}=\Lambda _{j}-\Lambda _{j-1}\text{\ \ and \ \ }\pi (s)=\left( \frac{j-2%
}{2}-\widetilde{\Lambda }_{j-1}\right) s.
\end{equation}
This condition leads to writting
\begin{equation}
\tau (s)=-2(1+\widetilde{\Lambda }_{j-1})s.
\end{equation}
Making use from Eqs. (29) and (33), the following expressions for $\lambda $
are obtained, respectively,

\begin{equation}
\lambda =\lambda _{n_{j}}=2n_{j}(1+\widetilde{\Lambda }%
_{j-1})+n_{j}(n_{j}-1),\text{ }j\in \lbrack 2,D-2],D>3
\end{equation}
\begin{equation}
\lambda =\Lambda _{j}-\Lambda _{j-1}-\widetilde{\Lambda }_{j-1}+\frac{j-2}{2}%
.
\end{equation}
Upon comparing Eqs. (42) and (43), we obtain

\begin{equation}
n_{j}=\widetilde{\ell }_{j}-\widetilde{\ell }_{j-1}.
\end{equation}
Additionally, using Eqs. (25)-(27) and (30)-(31), we obtain

\begin{equation}
\phi (s)=\left( 1-s^{2}\right) ^{\left( \frac{\widetilde{\Lambda }_{j-1}}{2}-%
\frac{(j-2)}{4}\right) },\text{ }\rho (s)=\left( 1-s^{2}\right) ^{\widetilde{%
\Lambda }_{j-1}},\text{ }j\in \lbrack 2,D-2],D>3.
\end{equation}
Besides, we substitute the wight function $\rho (s)$ given in Eq. (45) into
the Rodrigues relation Eq. (30) to obtain one of the wavefunctions in the
form
\begin{equation}
y_{n_{j}}(s)=A_{n_{j}}\left( 1-s^{2}\right) ^{-\widetilde{\Lambda }_{j-1}}%
\frac{d^{n}}{ds^{n}}\left( 1-s^{2}\right) ^{n_{j}+\widetilde{\Lambda }%
_{j-1}},
\end{equation}
where $A_{n_{j}}$ is the normaliation factor. Finally the angular
wavefunction is
\begin{equation}
H_{n_{j}}(\theta _{j})=N_{n_{j}}\left( \sin \theta _{j}\right) ^{^{\left(
\widetilde{\Lambda }_{j-1}-\frac{(j-2)}{2}\right) }}P_{n_{j}}^{(\widetilde{%
\Lambda }_{j-1},\widetilde{\Lambda }_{j-1})}(\cos \theta _{j}),\text{ }j\in
\lbrack 2,D-2],\text{ }D>3
\end{equation}
with the normalization factor
\begin{equation}
N_{n_{j}}=\sqrt{\frac{\left( 2\ell ^{\prime }+1\right) \Gamma (\ell ^{\prime
}-m^{\prime })}{2\Gamma (\ell ^{\prime }+m^{\prime })}}=\sqrt{\frac{\left(
2n_{j}+2\widetilde{\ell }_{j-1}+j-1\right) n_{j}!}{2\Gamma \left( n_{j}+2%
\widetilde{\ell }_{j-1}+j-2\right) }},\text{ \ }j\in \lbrack 2,D-2],\text{ }%
D>3.
\end{equation}
Likewise, in solving Eq. (35), we introduce a new variable $s=\cos \theta
_{D-1}.$ Thus, we can also rearrange it as the universal associated-Legendre
differential equation

\begin{equation}
\frac{d^{2}H(s)}{ds^{2}}-\frac{(D-1)s}{1-s^{2}}\frac{dH(s)}{ds}+\frac{\nu
^{\prime }(1-s^{2})-\Lambda _{D-2}^{\prime }}{(1-s^{2})^{2}}H(s)=0,
\end{equation}
where

\begin{equation}
\nu ^{\prime }=\ell ^{\prime }(\ell ^{\prime }+D-2)=\widetilde{\ell }(%
\widetilde{\ell }+D-2)+\beta \alpha _{2}^{2}\text{ \ \ and \ \ \ }\Lambda
_{D-2}^{\prime }{}=\Lambda _{D-2}+\beta \alpha _{2}^{2}.
\end{equation}
Equation (49) has been recently solved in $2D$ and $3D$ by the NU method in
\cite{22,25,30}. However, the aim in this subsection is to solve it in $D$%
-dimensions. Upon letting $D=3,$ we can readily restore $3D$ solution given
in \cite{30}. By comparing Eqs. (49) and (24), the corresponding polynomials are
obtained
\begin{equation}
\widetilde{\tau }(s)=-(D-1)s,\text{ \ \ \ }\sigma (s)=1-s^{2},\text{ \ \ }%
\widetilde{\sigma }(s)=-\nu ^{\prime }s^{2}+\nu ^{\prime }-\Lambda
_{D-2}^{\prime }.
\end{equation}
Inserting the above expressions into Eq. (32) and taking $\sigma ^{\prime
}(s)=-2s$, one obtains the following function:

\begin{equation}
\pi (s)=\frac{(D-3)}{2}s\pm \sqrt{\left[ \left( \frac{D-3}{2}\right)
^{2}+\nu ^{\prime }-k\right] s^{2}+k-\nu ^{\prime }+\Lambda _{D-2}^{\prime }}%
.
\end{equation}
Following the method, the polynomial $\pi (s)$ is found to have the
following four possible values:
\begin{equation}
\pi (s)=\left\{
\begin{array}{cc}
\left( \frac{D-3}{2}+\widetilde{\Lambda }_{D-2}\right) s & \text{\ for }%
k_{1}=\nu ^{\prime }-\Lambda _{D-2}^{\prime }, \\
\left( \frac{D-3}{2}-\widetilde{\Lambda }_{D-2}\right) s & \text{\ for }%
k_{1}=\nu ^{\prime }-\Lambda _{D-2}^{\prime }, \\
\frac{(D-3)}{2}s+\widetilde{\Lambda }_{D-2} & \text{\ for }k_{2}=\nu
^{\prime }+\left( \frac{D-3}{2}\right) ^{2}, \\
\frac{(D-3)}{2}s-\widetilde{\Lambda }_{D-2} & \text{\ for }k_{2}=\nu
^{\prime }+\left( \frac{D-3}{2}\right) ^{2},
\end{array}
\right.
\end{equation}
where $\widetilde{\Lambda }_{D-2}=\sqrt{\left( \widetilde{\ell }_{D-2}+\frac{%
D-3}{2}\right) ^{2}+\beta \alpha _{2}^{2}}.$ Imposing the condition $\tau
^{\prime }(s)<0$ for Eq. (28), one selects the following physically valid
solutions with $\tau ^{\prime }=\tau ^{\prime }(\widetilde{\ell }_{D-2});$
that is, a function of the angular momentum,\footnote{%
The physical significance of this choice of parameters is that the
eigenvalue and the eigenfunction equations can be directly reduced to the
three-dimensional form in Ref. \cite{30}.}

\begin{equation}
k_{1}=\nu ^{\prime }-\Lambda _{D-2}^{\prime }\text{ \ \ and \ \ }\pi
(s)=\left( \frac{D-3}{2}-\widetilde{\Lambda }_{D-2}\right) s,
\end{equation}
which yields from Eq. (28) that
\begin{equation}
\tau (s)=-2(1+\widetilde{\Lambda }_{D-2})s.
\end{equation}
Making use from Eqs. (29) and (33), the following expressions for $\lambda $
are obtained, respectively,

\begin{equation}
\lambda =\lambda _{n_{D-1}}=2n_{D-1}(1+\widetilde{\Lambda }%
_{D-2})+n_{D-1}(n_{D-1}-1),
\end{equation}
\begin{equation}
\lambda =\nu ^{\prime }-\Lambda _{D-2}^{\prime }-\widetilde{\Lambda }_{D-2}+%
\frac{D-3}{2}.
\end{equation}
We compare Eqs. (56) and (57) and from the definition $\nu ^{\prime }=\ell
^{\prime }(\ell ^{\prime }+D-2),$ the new angular momentum\ $\ell ^{\prime },%
\widetilde{\ell }_{D-2}$ and $\widetilde{\ell }$ values are obtained as

\[
\widetilde{\ell }=-\frac{\left( D-2\right) }{2}+\sqrt{\left( n_{j}+m^{\prime
}+\frac{1}{2}\right) ^{2}-\alpha _{2}^{2}\beta },
\]
\[
\ell ^{\prime }=-\frac{1}{2}+\sqrt{\left( \widetilde{\ell }+\frac{D-2}{2}%
\right) ^{2}+\beta \alpha _{2}^{2}}=n_{D-1}+m^{\prime },
\]
\begin{equation}
m^{\prime }=\sqrt{\left( \widetilde{\ell }_{D-2}+\frac{D-3}{2}\right)
^{2}+\alpha _{2}^{2}\beta },
\end{equation}
where $\widetilde{\ell }=\widetilde{\ell }_{D-1},$ which can be easily
reduced to the well-known definition
\begin{equation}
\ell ^{\prime }=n+\sqrt{m^{2}+\alpha _{2}^{2}\beta },
\end{equation}
where $n_{D-1}=n,$ $\widetilde{\ell }_{D-2}=m,$ $\widetilde{\Lambda }%
_{D-2}=m^{\prime \text{ }}$in $3D$ space [30]. Using Eqs. (25-(27) and
(30)-(31), we obtain

\begin{equation}
\phi (s)=\left( 1-s^{2}\right) ^{(2\widetilde{\Lambda }_{D-2}+3-D)/4},\text{
}\rho (s)=\left( 1-s^{2}\right) ^{\widetilde{\Lambda }_{D-2}}.
\end{equation}
Besides, we substitute the wight function $\rho (s)$ given in Eq. (60) into
the Rodrigues relation (30) and obtain one of the wavefunctions in the form
\begin{equation}
y_{n_{D-1}}(s)=B_{n_{D-1}}\left( 1-s^{2}\right) ^{-\widetilde{\Lambda }%
_{D-2}}\frac{d^{n_{D-1}}}{ds^{n_{D-1}}}\left( 1-s^{2}\right) ^{n_{D-1}+%
\widetilde{\Lambda }_{D-2}},
\end{equation}
where $B_{n_{D-1}}$ is the normalization factor. Finally the angular
wavefunction is
\begin{equation}
H_{n_{D-1}}(\theta _{D-1})=N_{n_{D-1}}\left( \sin \theta _{D-1}\right) ^{%
\widetilde{\Lambda }_{D-2}-\frac{(D-3)}{2}}P_{n_{D-1}}^{(\widetilde{\Lambda }%
_{D-2},\widetilde{\Lambda }_{D-2})}(\cos \theta _{D-1}),
\end{equation}
where the normalization factor
\begin{equation}
N_{n_{D-1}}=\sqrt{\frac{\left( 2n_{D-1}+2m^{\prime }+1\right) n_{D-1}!}{%
2\Gamma \left( n_{D-1}+2m^{\prime }\right) }},
\end{equation}
where $m^{\prime }$ is given in Eq. (58).

\subsection{The eigenvalues and eigenfunctions of the radial equation}

We seek to present the exact bound-state solutions, i.e., the energy spectra
and radial wave function $R_{\widetilde{\ell }}(r)$ of the Klein-Gordon
equation in Eq. (22), by simply writting it in the following simple form
\cite{22,25,41}:

\begin{equation}
\frac{d^{2}g(r)}{dr^{2}}-\left[ \frac{(M-1)(M-3)}{4r^{2}}+\alpha
_{2}^{2}\left( Ar^{2}+\frac{B}{r^{2}}+C\right) +\alpha _{1}^{2}\alpha
_{2}^{2}\right] g(r)=0,
\end{equation}
where
\begin{equation}
M=D+2\widetilde{\ell },
\end{equation}
with $\widetilde{\ell }$ is given in (58). It is worthwhile to note that for
bound states, we require that the wavefunction $g(r)$ must satisfy the
boundary condition that $g(r)$ becomes zero where $r\rightarrow \infty $,
and $g(r)$ is finite at $r=0.$ Further, applying the following simple
transformation of variables, $s=r^{2},$ and making some algebraic
manipulations, we may rewrite Eq. (64) in the standard form,
\begin{equation}
\frac{d^{2}g(s)}{ds^{2}}+\frac{1}{2s}\frac{dg(s)}{ds}+\frac{1}{(2s)^{2}}%
\left[ -\alpha ^{2}s^{2}-\varepsilon ^{2}s-\gamma ^{2}\right] g(s)=0,
\end{equation}
with the following definitions
\begin{equation}
\varepsilon ^{2}=\alpha _{2}^{2}(\alpha _{1}^{2}+C),\text{ 4}\gamma
^{2}=(M-1)(M-3)+4B\alpha _{2}^{2},\text{ }\alpha ^{2}=A\alpha _{2}^{2}.
\end{equation}
Comparing Eq. (66) with Eq. (24), gives the following expressions:

\begin{equation}
\widetilde{\tau }(s)=1,\text{ }\sigma (s)=2s,\text{ }\widetilde{\sigma }%
(r)=-\alpha ^{2}s^{2}-\varepsilon ^{2}s-\gamma ^{2}.
\end{equation}
Substituting the above expressions into Eq. (32) gives

\begin{equation}
\pi (s)=\frac{1}{2}\pm \frac{1}{2}\sqrt{4\alpha ^{2}s^{2}+4(\varepsilon
^{2}+2k)s+4\gamma ^{2}+1}.
\end{equation}
Therefore, we can determine the constant $k$ by using the condition that the
discriminant of the square root is zero, that is
\begin{equation}
k=-\frac{\varepsilon ^{2}}{2}\pm \frac{\alpha }{2}\sqrt{4\gamma ^{2}+1},%
\text{ }4\gamma ^{2}+1=(D+2j-2)^{2}+4B\alpha _{2}^{2}.
\end{equation}
In view of that, we arrive at the following four possible functions of $\pi
(s):$%
\begin{equation}
\pi (s)=\left\{
\begin{array}{cc}
\frac{1}{2}+\left[ \alpha s+\frac{1}{2}\sqrt{4\gamma ^{2}+1}\right] & \text{%
\ for }k_{1}=-\frac{\varepsilon ^{2}}{2}+\frac{\alpha }{2}\sqrt{4\gamma
^{2}+1}, \\
\frac{1}{2}-\left[ \alpha s+\frac{1}{2}\sqrt{4\gamma ^{2}+1}\right] & \text{%
\ for }k_{1}=-\frac{\varepsilon ^{2}}{2}+\frac{\alpha }{2}\sqrt{4\gamma
^{2}+1}, \\
\frac{1}{2}+\left[ \alpha s-\frac{1}{2}\sqrt{4\gamma ^{2}+1}\right] & \text{%
\ for }k_{2}=-\frac{\varepsilon ^{2}}{2}-\frac{\alpha }{2}\sqrt{4\gamma
^{2}+1}, \\
\frac{1}{2}-\left[ \alpha s-\frac{1}{2}\sqrt{4\gamma ^{2}+1}\right] & \text{%
\ for }k_{2}=-\frac{\varepsilon ^{2}}{2}-\frac{\alpha }{2}\sqrt{4\gamma
^{2}+1}.
\end{array}
\right.
\end{equation}
The correct value of $\pi (s)$ is chosen such that the function $\tau (s)$
given by Eq. (28) will have negative derivative \cite{27}. So we can select the
physical values to be

\begin{equation}
k=-\frac{\varepsilon ^{2}}{2}-\frac{\alpha }{2}\sqrt{4\gamma ^{2}+1}\text{ \
\ and \ \ }\pi (s)=\frac{1}{2}-\left[ \alpha s-\frac{1}{2}\sqrt{4\gamma
^{2}+1}\right] ,
\end{equation}
which yield
\begin{equation}
\tau (s)=-2\alpha s+2+\sqrt{4\gamma ^{2}+1},\text{ }\tau ^{\prime
}(s)=-2\alpha <0.
\end{equation}
Using Eqs. (29) and (33), the following expressions for $\lambda $ are
obtained, respectively,

\begin{equation}
\lambda =\lambda _{n}=2\alpha n,\text{ }n=0,1,2,...
\end{equation}
\begin{equation}
\lambda =-\frac{\varepsilon ^{2}}{2}-\frac{\alpha }{2}(2+\sqrt{4\gamma ^{2}+1%
}).
\end{equation}
So we can obtain the energy levels of the Klein-Gordon from the following
relation:
\begin{equation}
-\sqrt{A}\left[ 4n+2+\sqrt{\left( D+2\widetilde{\ell }-2\right) ^{2}+4(\mu
+E_{R})B}\right] =\sqrt{\mu +E_{R}}\left( \mu -E_{R}+C\right) ,
\end{equation}
and hence for the pseudoharmonic plus the new ring-shaped potential, it
becomes

\begin{equation}
-\sqrt{a_{0}}\left[ 4n+2+\sqrt{\left( D+2\widetilde{\ell }-2\right)
^{2}+4a_{0}r_{0}^{2}(\mu +E_{R})}\right] =r_{0}\sqrt{\mu +E_{R}}\left( \mu
-E_{R}-2a_{0}\right) ,
\end{equation}
where $\widetilde{\ell }$ is given in Eq. (58). The energy $E_{R}$ is
defined implicitly by Eq. (77) which is a rather complicated transcendental
equation having many solutions for a given values of $n$ and $\widetilde{%
\ell }.$

For completeness, we find that it is necessary to consider the solution for
the central harmonic oscillator potential, $V(r)=\frac{1}{2}k^{2}r^{2}$
[33]. Therefore, applying the parameters transformation for this potential
as:: $A=\frac{1}{2}k^{2},$ $B=C=\beta =0$ and $\widetilde{\ell }=\ell ,$ the
non central potential in (4) turns into the harmonic oscillator with Klein
Gordon solution for the energy spectra as
\begin{equation}
(\mu +E_{R})\left( \mu -E_{R}\right) ^{2}=\frac{k^{2}}{2}\left[ 4n+2\ell +D%
\right] ^{2},\text{ }n,\ell =0,1,2,....
\end{equation}
On the other hand, in the nonrelativistic limit, applying the following
appropriate transformation: $\mu +E_{R}\rightarrow 2\mu ,$ $\mu
-E_{R}\rightarrow -$ $E_{NR},$ $\widetilde{\ell }\rightarrow \ell $ to Eq.
(78) yields \cite{33,34}
\begin{equation}
E_{NR}=\frac{k}{\sqrt{\mu }}\left[ 2n+\ell +\frac{D}{2}\right] ,\text{ \ }%
n,\ell =0,1,2,...
\end{equation}
In addition, from Eq. (76), we obtain the solution for the pseudoharmonic
potential $(\beta =0$ case$)$ in the $3D$-Schr\"{o}dinger equation as \cite{34}

\begin{equation}
E_{NR}=-2a_{0}+\sqrt{\frac{2a_{0}}{\mu r_{0}^{2}}}\left[ 2n+1+\sqrt{\left(
\ell +\frac{1}{2}\right) ^{2}+2\mu a_{0}r_{0}^{2}}\right] ,
\end{equation}
and for the ring-shaped pseudoharmonic potential $(\beta \neq 0$ case$)$
\cite{25}:

\begin{equation}
E_{NR}=-2a_{0}+\sqrt{\frac{2a_{0}}{\mu r_{0}^{2}}}\left[ 2n+1+\sqrt{\left(
\widetilde{n}+\sqrt{m^{2}+2\mu \beta }+\frac{1}{2}\right) ^{2}+2\mu \left(
a_{0}r_{0}^{2}-\beta \right) }\right] ,
\end{equation}
where $\widetilde{n}$ and $m$ are two constants coming from the solution of
the angular wave equation with $\ell =\widetilde{n}+\sqrt{m^{2}+2\mu \beta }%
. $

Further, inserting the values of $\sigma (s),\pi (s)$ and $\tau (s)$ in Eqs
(37), (40) and (41) into Eqs. (27) and (31), we obtain the wavefunctions
\begin{equation}
\phi (s)=s^{(\zeta +1)/4}e^{-\alpha s/2},
\end{equation}

\begin{equation}
\rho (s)=s^{\zeta /2}e^{-\alpha s},
\end{equation}
where $\zeta =\sqrt{4\gamma ^{2}+1}.$ Further, from Eq. (30), we obtain

\begin{equation}
y_{n\widetilde{\ell }}(s)=B_{n\widetilde{\ell }}e^{\alpha s}s^{-\zeta /2}%
\frac{d^{n}}{ds^{n}}\left[ e^{-\alpha s}s^{\zeta /2}\right] =B_{n\widetilde{%
\ell }}L_{n}^{(\Lambda +1/2)}(\alpha s),
\end{equation}
where $2\Lambda +1=\zeta $ and consequently the wave function $g(s)$ can be
expressed in terms of the generalized Laguerre polynomials as

\begin{equation}
g(s)=C_{n\widetilde{\ell }}s^{(\Lambda +1)/2}e^{-\alpha s/2}L_{n}^{(\Lambda
+1/2)}(\alpha s),
\end{equation}
where for the ring-shaped pseudoharmonic potential we have
\begin{equation}
\zeta =\sqrt{\left( D+2\widetilde{\ell }-2\right) ^{2}+4a_{0}r_{0}^{2}(\mu
+E_{R})},\text{ }\alpha =\frac{\sqrt{a_{0}(\mu +E_{R})}}{r_{0}}.
\end{equation}
Finally, the radial wave functions of the Klein-Gordon equation are obtained
from Eqs. (9) and (85) as
\begin{equation}
R\widetilde{\ell }(r)=C_{n\widetilde{\ell }}r^{\Lambda +1-(D-1)/2}\exp
(-\alpha r^{2}/2)L_{n}^{(\Lambda +1/2)}(\alpha r^{2}),
\end{equation}
where \cite{46,47}

\begin{equation}
C_{n\widetilde{\ell }}=\sqrt{\frac{2\alpha ^{\Lambda +3/2}n!}{\Gamma \left(
\Lambda +n+3/2\right) }},
\end{equation}
and we can finally obtain the re-normalized total wavefunctions

\[
\psi _{\ell _{1}\cdots \ell _{D-2}}^{(\ell )}({\bf x})=\frac{1}{2^{m^{\prime
}}(\widetilde{n}+m^{\prime })!}\sqrt{\frac{\alpha ^{\Lambda +3/2}n!}{\pi
\Gamma \left( \Lambda +n+3/2\right) }}
\]
\[
\times r^{\Lambda +1-(D-1)/2}\exp (-\alpha r^{2}/2)L_{n}^{(\Lambda
+1/2)}(\alpha r^{2})
\]
\[
\exp (\pm im\varphi )\prod\limits_{j=2}^{D-2}\sqrt{\frac{\left( 2n_{j}+2%
\widetilde{\ell }_{j-1}+j-1\right) n_{j}!}{2\Gamma \left( n_{j}+2\widetilde{%
\ell }_{j-1}+j-2\right) }}\left( \sin \theta _{j}\right) ^{\left( \widetilde{%
\Lambda }_{j-1}-\frac{(j-2)}{2}\right) }P_{n_{j}}^{(\widetilde{\Lambda }%
_{j-1},\widetilde{\Lambda }_{j-1})}(\cos \theta _{j})
\]
\begin{equation}
\sqrt{\frac{\left( 2n_{D-1}+2m^{\prime }+1\right) n_{D-1}!}{2\Gamma \left(
n_{D-1}+2m^{\prime }\right) }}\sin (\theta _{D-1})^{\widetilde{\Lambda }%
_{D-2}-(D-3)/2}P_{n_{D-1}}^{(\widetilde{\Lambda }_{D-2},\widetilde{\Lambda }%
_{D-2})}(\cos \theta _{D-1}),
\end{equation}
where $\Lambda =\frac{1}{2}(\zeta -1)$.

\section{Conclusions}

\label{C}We have calculated the exact bound-state energy\ eigenvalues and
the corresponding eigenfunctions of the relativistic spin-$0$ particle in
the $D$-dimensional Klein-Gordon equation with equal scalar and vector
ring-shaped pseudoharmonic potential using the NU method. The analytical
expressions for the total energy levels and eigenfunctions of this system
can be reduced to their well-known $3D$ Schr\"{o}dinger equation. Further,
the noncentral potentials treated in Ref. [20] can be introduced as
perturbation to the pseudoharmonic potential by adjusting the strength of
the coupling constant $\beta $ in terms of $a_{0},$ which is the coupling
constant of the pseudoharmonic potential. The relativistic energy $E_{R}$
defined implicitly by Eq. (76) is rather complicated transcendental equation
and it has many solutions for any arbitrarily given values of $n$ and $%
\widetilde{\ell }.$ \ Additionally, the radial and polar angle wave
functions of the Klein-Gordon equation are found in terms of Laguerre and
Jacobi polynomials, respectively. The method presented in this paper is
general and worth extending to the solution of other interaction problems.
This method is very simple and useful in solving other complicated systems
analytically without given a restriction conditions on the solution of some
quantum systems as the case in the other models. Therefore, we have seen
that for the nonrelativistic model, the exact energy spectra can be obtained
either by solving the Schr\"{o}dinger equation in Eq. (11) (cf. Eq. (48) in
Ref. \cite{25}) or rather by applying appropriate transformation to the
relativistic solution. Finally, we emphasize that the exact bound-state
spectra obtained in this work might have some interesting applications in
different branches like atomic and molecular physics and quantum chemistry.
They describe the molecular structures and interactions in diatomic
molecules \cite{48}.

\acknowledgments This research was partially supported by the Scientific and
Technical Research Council of Turkey. 


\bigskip

\begin{thebibliography}{99}
\bibitem{1}  T. Y. Wu and W. Y. Pauchy Hwang, Relativistic Quantum Mechanics
and Quantum Fields (World Scientific, Singapore, 1991).

\bibitem{2}  W. Greiner, Relativistic Quantum Mechanics: Wave Equations, 3rd
edn (springer, Berlin, 2000).

\bibitem{3}  A. D. Alhaidari, Phys. Rev. Lett. 87 (2001) 210405; 88 (2002)
189901.

\bibitem{4}  G. Chen, Mod. Phys. Lett. A 19 (2004) 2009; J. -Y. Guo, J. Meng
and F. -X. Xu, Chin. Phys. Lett. 20 (2003) 602; A. D. Alhaidari, J.
Phys. A:Math. Gen. 34 (2001) 9827; 35 (2002) 6207; M. \c{S}im\c{s}ek and H. E\u{g}%
rifes, J. Phys. A: Math. Gen. 37 (2004) 4379.

\bibitem{5}  J. -Y. Guo, X. -Z. Fang and F. -X. Xu, Phys. Rev. A 66 (2002)
062105; C. Berkdemir, A. Berkdemir and R. Sever, J. Phys. A: Math. Gen. 39
(2006) 13455.

\bibitem{6}  G. Chen, Acta Phys. Sinica 50 (2001) 1651; \"{O}. Ye\c{s}ilta%
\c{s}, Phys. Scr. 75 (2007) 41.

\bibitem{7}  G. Chen and Z.M. Lou, Acta Phys. Sinica 52 (2003) 1071.

\bibitem{8}  G. Chen, Z. D. Chen and Z. M. Lou, Chin. Phys. 13 (2004) 279.

\bibitem{9}  W. C. Qiang, Chin. Phys. 12 (2003) 136.

\bibitem{10}  W. C. Qiang, Chin. Phys. 13 (2004) 571.

\bibitem{11}  G. Chen, Phys. Lett. A 328 (2004) 116; Y. F. Diao, L. Z. Yi and
C. S. Jia, Phys. Lett. A 332 (2004) 157.

\bibitem{12}  L. Z. Yi {\it et al}, Phys. Lett. A 333 (2004) 212.

\bibitem{13}  X. Q. Zhao, C. S. Jia and Q. B. Yang, Phys. Lett. A 337 (2005)
189.

\bibitem{14}  A. D. Alhaidari, H. Bahlouli and A. Al-Hasan, Phys. Lett. A 349
(2006) 87.

\bibitem{15}  G. Chen, Phys. Lett. A 339 (2005) 300.

\bibitem{16}  A. de Souza Dutra and G. Chen, Phys. Lett. A 349 (2006) 297.

\bibitem{17}  F. Dominguez-Adame, Phys. Lett. A 136 (1989) 175.

\bibitem{18}  A. S. de Castro, Phys. Lett. A 338 (2005) 81.

\bibitem{19}  G. Chen, Acta Phys. Sinica 53 (2004) 680; G. Chen and D. F.
Zhao, Acta Phys. Sinica 52 (2003) 2954.

\bibitem{20}  S. M. Ikhdair and R. Sever, Int. J. Theor. Phys. 46 (10) (2007)
2384.

\bibitem{21}  M. Kibler and P. Winternitz, J. Phys. A 20 (1987) 4097.

\bibitem{22}  S. M. Ikhdair and R. Sever{\it ,} Int. J. Mod. Phys. C 18 (10)
(2007) 1571; preprint quant-ph/0703008; to appear in Int. J. Mod.
Phys. C.

\bibitem{23}  H. Hartmann and D. Schuch, Int. J. Quantum Chem. 18 (1980) 125.

\bibitem{24}  C. Quesne, J. Phys. A 21 (1988) 3093.

\bibitem{25}  S. M. Ikhdair and R. Sever{\it ,} preprint quant-ph/0703131, to
appear in Cent. E. J. Phys.

\bibitem{26}  F. Yasuk, A. Durmu\c{s} and I. Boztosun, J. Math. Phys. 47
(2006) 082302.

\bibitem{27}  A. F. Nikiforov and V. B. Uvarov, Special Functions of
Mathematical Physics (Birkhauser, Bassel, 1988).

\bibitem{28}  C. Berkdemir, Am. J. Phys. 75 (2007) 81.

\bibitem{29}  C. Y. Chen and S.H. Dong, Phys. Lett. A 335 (2005) 374.

\bibitem{30}  Y. F. Cheng and T. Q. Dai, Phys. Scr. 75 (2007) 274.

\bibitem{31}  C. Berkdemir, A. Berkdemir and J. G. Han, Chem. Phys. Lett. 417
(2006) 326.

\bibitem{32}  S. M. Ikhdair and R. Sever{\it ,} arXiv: 0704.0573 to appear in
the Cent. E. J. Phys.

\bibitem{33}  M. Sage and J. Goodisman, Am. J. Phys. 53 (1985) 350.

\bibitem{34}  S. M. Ikhdair and R. Sever, J. Mol. Struct.-Theochem 806 (2007)
155; Cent. E. J. Phys. 5 (4) (2007) 516; preprint quant-ph/0702052
to appear in the Cent. E. J. Phys.; doi:10.1016/j.theochem
2007.12.044, to appear in J. Mol. Struct.-Theochem.

\bibitem{35}  L. -Y. Wang, X. -Y. Gu, \ Z. -Q. Ma and S. -H. Dong, Found.
Phys. Lett..15 (2002) 569; S. -H. Dong, App. Math. Lett. 16
(2003)199.

\bibitem{36}  J. D. Louck and W. H. Shaffer, J. Mol. Spec. 4 (1960) 285; D.
Louck, J. Mol. Spec. 4 (1960) 298; J. D. Louck, J. Mol. Spec.4
(1960) 334.

\bibitem{37}  J. D. Louck, Theory of Angular Momentum in D-Dimensional Space,
Los Alamos Scientific Laboratory monograph LA-2451, LASL, Los
Alamos, 1960.

\bibitem{38}  J. D. Louck and H. W. Galbraith, Rev. Mod. Phys. 48 (1976) 69

\bibitem{39}  A. Chatterjee, Phys. Rep. 186 (1990) 249.

\bibitem{40}  A. Erd\'{e}lyi, Higher Transcendental Functions, Vol. 2, McGraw
Hill, New York, 1953.

\bibitem{41}  S. M. Ikhdair and R. Sever{\it ,} Z. Phys. C 56 (1992) 155; C 58
(1993) 153; D 28 (1993) 1; Hadronic J. 15 (1992) 389; Int. J. Mod.
Phys. A 18 (2003) 4215; A 19 (2004) 1771; A 20 (2005) 4035; A 20
(2005) 6509; A 21 (2006) 2191; A 21 (2006) 3989; A 21 (2006) 6699;
Int. J. Mod. Phys. E (in press) (preprint hep-ph/0504176); S.
Ikhdair {\it et al,} Tr. J. Phys. 16 (1992) 510; 17 (1993) 474.

\bibitem{42}  S. M. Ikhdair, preprint quant-ph/0703042, to appear in the
Chinese J. Phys.

\bibitem{43}  S. M. Ikhdair and R. Sever, Int. J. Theor. Phys. 46 (6) (2007)
1643; J. Math. Chem. 41 (2007) 329; 41 (2007) 343; 42 (3) (2007) 461.

\bibitem{44}  S. M. Ikhdair and R. Sever, Ann. Phys. (Leipzig) 16 (3) (2007)
218.

\bibitem{45}  S. M. Ikhdair and R. Sever, preprint quant-ph/0605045 to appear
in the Int. J. Mod. Phys. E.

\bibitem{46}  G. Sezgo, Orthogonal Polynomials (American Mathematical Society,
New York, 1939).

\bibitem{47}  N. N. Lebedev, Special Functions and Their Applications
(Prentice-Hall, Englewood Cliffs, NJ, 1965).

\bibitem{48}  R. J. Le Roy and R. B. Bernstein, J. Chem. Phys. 52 (1970) 3869.
\end{thebibliography}
\end{document}